%% file: siggraph_asia_25.tex
\documentclass[acmtog]{acmart}
\acmSubmissionID{1660}

\input{preamble}

\acmJournal{TOG}

\copyrightyear{2025}
\acmYear{2025}
\setcopyright{cc}
\setcctype{by}
\acmConference[SA Conference Papers '25]{SIGGRAPH Asia 2025 Conference Papers}{December 15--18, 2025}{Hong Kong, Hong Kong}
\acmBooktitle{SIGGRAPH Asia 2025 Conference Papers (SA Conference Papers '25), December 15--18, 2025, Hong Kong, Hong Kong}\acmDOI{10.1145/3757377.3763904}
\acmISBN{979-8-4007-2137-3/2025/12}

\begin{document}

\title{\textcolor{title_purple}{In-2-4D}: \textcolor{title_purple}{In}betweening from \textcolor{title_purple}{Two} Single-View Images to \textcolor{title_purple}{4D} Generation}

\renewcommand{\shortauthors}{}

\author{Sauradip Nag}
\affiliation{%
 \institution{Simon Fraser University}
 \country{Canada}
}
\email{snag@sfu.ca}

\author{Daniel Cohen-Or}
\affiliation{%
 \institution{Tel Aviv University}
 \country{Israel}
}
\email{dcor@gmail.com}

\author{Hao Zhang}
\affiliation{%
 \institution{Simon Fraser University}
 \country{Canada}
}
\email{haoz@sfu.ca}

\author{Ali Mahdavi-Amiri}
\affiliation{%
 \institution{Simon Fraser University}
 \country{Canada}
}
\email{amahdavi@sfu.ca}

\begin{abstract}
\input{main/00_abs}
\end{abstract}


\maketitle

\input{main/01_intro}

\input{main/02_lit}
\input{main/03_model}
\input{main/04_exp}
\input{main/05_conclusion}


\bibliographystyle{ACM-Reference-Format}
\bibliography{references}

\begin{figure*}[ht]
\centering
    \includegraphics[width=1\textwidth]{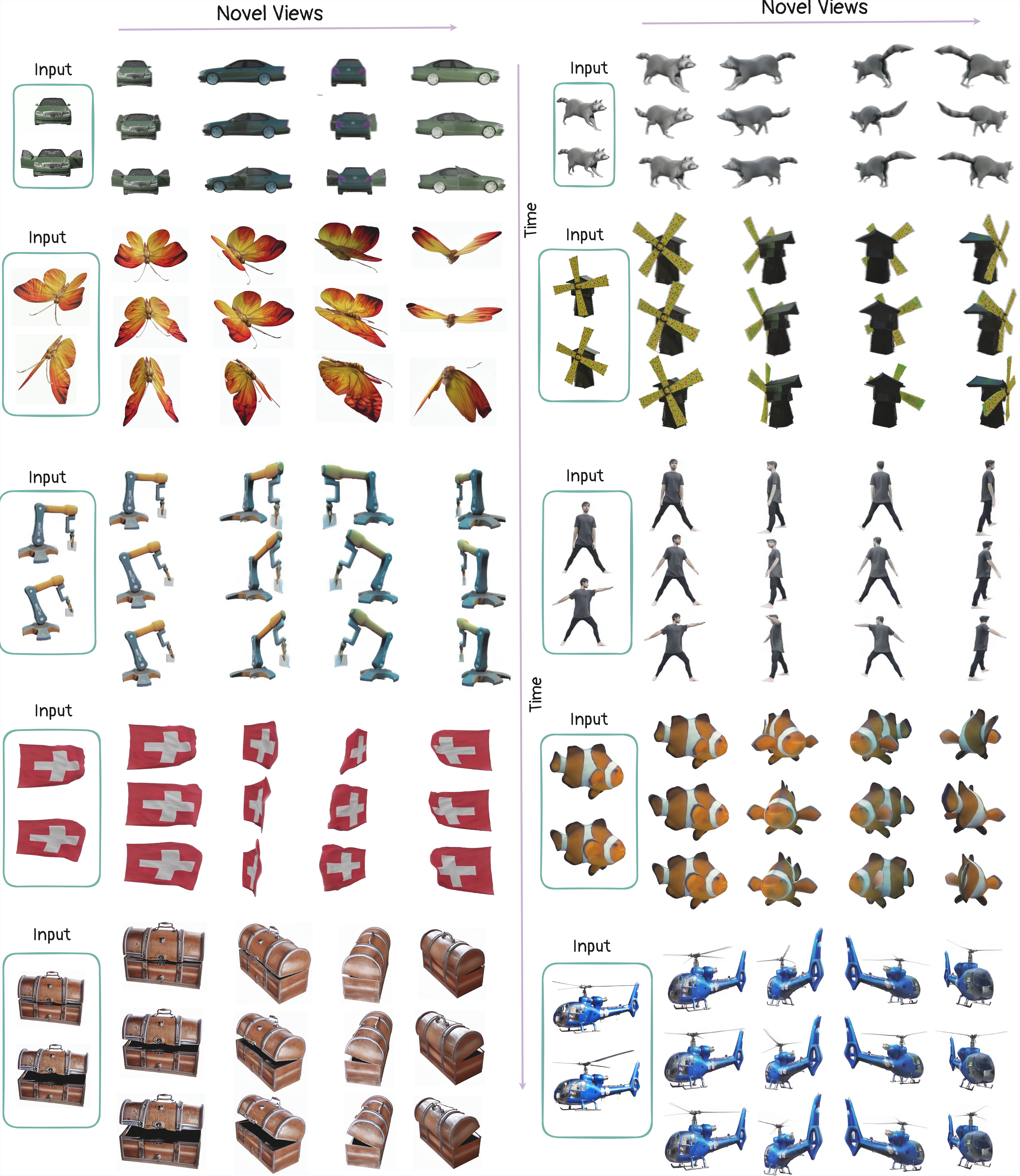}
    \caption{\cready{\textbf{Results Gallery.} We illustrate our method's performance on a wide variety of objects including furnitures, animals, humans, vehicles and daily life objects. Our model produces consistent textures and preserves the geometry in novel views without any direct 3D supervision.}}
    \label{fig:qual-gallery}
\end{figure*}

\begin{figure*}[ht]
\centering
    \includegraphics[width=1\textwidth]{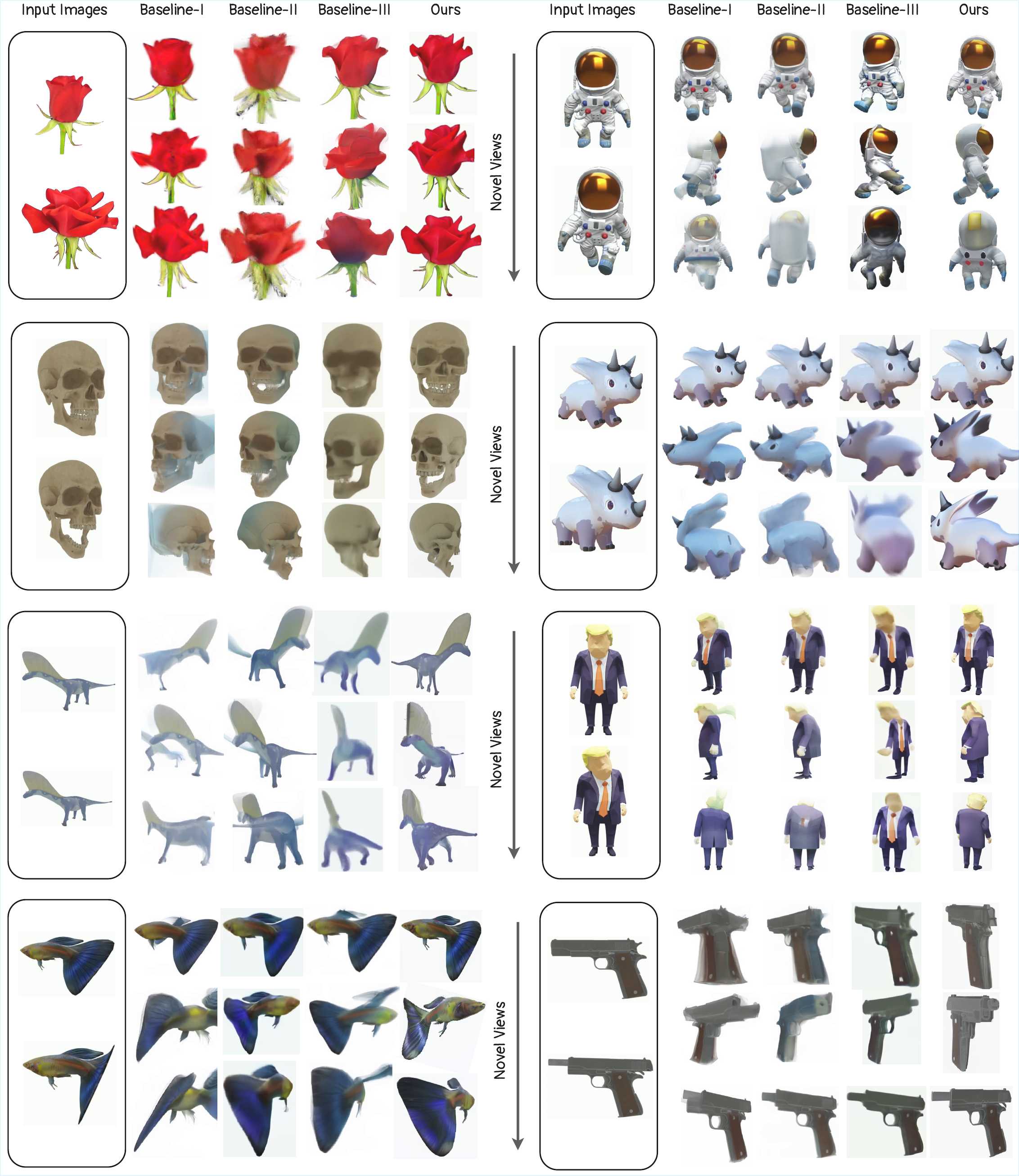}
    \caption{\cready{\textbf{Qualitative Comparison on Consistent4D benchmark.} Having only the first and last frames of a motion, we are able to generate moving 3D objects that can be seen at different views. Objects seen from different view directions are still plausible although no direct supervision signal is available.}}
    \label{fig:qual-c4d}
\end{figure*}

\end{document}

%% file: preamble.tex
\usepackage[table]{xcolor}
\usepackage{multirow}
\usepackage{pifont}

\newcommand{\cmark}{\ding{51}}%
\newcommand{\xmark}{\ding{55}}%
\definecolor{iccvblue}{rgb}{0.21,0.49,0.74}

\usepackage{graphicx,amsmath,amsthm,booktabs,listings,hyperref,microtype,nicefrac,enumitem,bbm,amsfonts,tikz}
\usepackage[ruled,vlined]{algorithm2e}
\usepackage{multirow}
\usepackage{cutwin}
\usepackage{mdframed}
\newmdenv[linewidth=0.8pt,leftmargin=0pt,topline=false,rightline=false,bottomline=false]{assbox}

\SetAlFnt{\small}
\SetAlCapFnt{\small}
\SetAlCapNameFnt{\small}
\SetAlCapHSkip{0pt}
\definecolor{title_purple}{rgb}{0.65,0.1,0.65}
\newcommand{\method}[1]{{\bf In-2-4D}}

\definecolor{jointyellow}{RGB}{233, 184, 44}

\definecolor{title_purple}{rgb}{0.65,0.1,0.65}

\newcommand{\cready}[1]{\textcolor{black}{#1}}

\definecolor{darkg}{RGB}{0, 150, 0}
\newcommand{\rz}[1]{\textcolor{black}{#1}}
\newcommand{\rzz}[1]{\textcolor{black}{#1}}
\newcommand{\rzzz}[1]{\textcolor{black}{#1}}
\newcommand{\am}[1]{\textcolor{black}{#1}}
\newcommand{\saura}[1]{\textcolor{black}{#1}}
\newcommand{\snn}[1]{\textcolor{black}{#1}}

\definecolor{titlepurple}{RGB}{148, 103, 189}

%% file: main/00_abs.tex
We \rzzz{pose} a new problem, \textbf{In-2-4D}, for generative 4D (i.e., 3D + motion) inbetweening \rzzz{to interpolate} two {\em single-view\/} images. 
\rzzz{In contrast to video/4D generation from only text or a single image, our interpolative task can leverage more precise motion control to better constrain the generation.}
Given two monocular RGB images representing the start and end states of an object in motion, our goal is to generate and reconstruct the motion in 4D, 
\rzzz{without making assumptions on the object category, motion type, length, or complexity.}
To handle such arbitrary and diverse motions, we utilize a \rzzz{foundational} video interpolation model for motion prediction. However, large frame-to-frame motion \rzzz{gaps} can lead to ambiguous interpretations. 
To this end, we employ a {\em hierarchical\/} approach to identify keyframes that are visually close to the input states \rzzz{while exhibiting} significant motions, then generate smooth fragments between them. 
For each fragment, we construct a 3D representation of the keyframe using Gaussian Splatting (3DGS). 
The temporal frames within the fragment guide the motion, enabling their transformation into dynamic 3DGS through a deformation field. 
To improve temporal consistency and refine the 3D motion, we expand the self-attention of multi-view diffusion across timesteps and apply rigid transformation regularization. 
Finally, we merge the independently generated 3D motion segments by interpolating boundary deformation fields and optimizing them to align with the guiding video, ensuring smooth and flicker-free transitions. 
Through extensive qualitative and quantitive experiments as well as a user study, we demonstrate the effectiveness of our method and design choices. 

\noindent
\textbf{Project Page \& Source Code:} \href{https://in-2-4d.github.io/}{\text{https://in-2-4d.github.io/}}

%% file: main/01_intro.tex
\section{Introduction}
\label{sec:intro}

\begin{figure}[!t]
    \centering
    \includegraphics[width=\linewidth]{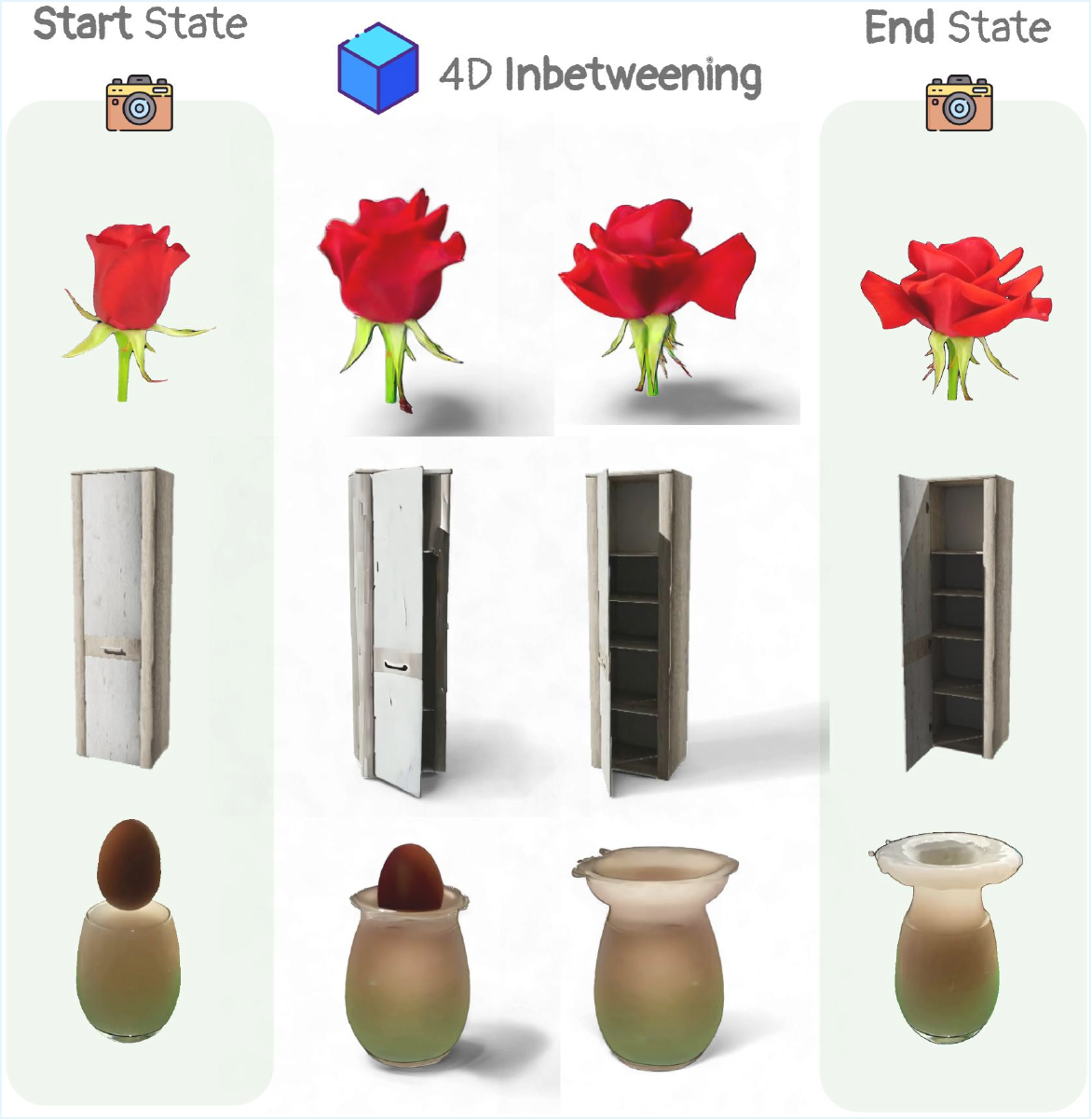}
    \captionof{figure}{\textbf{In-2-4D: 4D motion inbetweening from 2 single-view images,} \rzzz{a minimalistic input setup while still offering visual motion control.} Given two monocular RGB images of an object at two distinct motion states (start and end),
our method generates a smooth, natural, and seamless 4D (3D object + motion) interpolation between them.
\rzzz{We make no assumptions on the object category or motion type. Piecewise rigid, freeform, and long-range motions, even with topology changes (e.g., egg dropping in liquid causing a splash), are all allowed.}
}
    \label{fig:teaser}
\end{figure}%


Motion inbetweening is a classic animation problem. When generating motions of 3D objects, the typical input consists of
a 3D object in two distinct motion states, e.g., as in point cloud interpolation~\cite{peng2024papr,zheng2023neuralpci}. 
With significant progresses in 3D generative AI in recent years,
many recent attempts have been made on ``video-to-4D" \cite{zeng2024stag4d,wu2024sc4d,ren2023dreamgaussian4d,yin20234dgen,jiang2023consistent4d}, whose task is to ``lift'' an 
object captured in a video into the 3D space so its motion from the video can be viewed from all angles.
\rzzz{However, videos can be unrealistic (e.g., for very slow motions such as the flower blossoming in Fig.~\ref{fig:teaser}) or at least expensive (e.g., for very fast motions) to acquire.
In other application settings such as motion planning and visual storyboarding~\cite{zhong2025Sketch2Anim}, the motions produced are the outcomes to be {\em explored\/}; they were not acquirable in the first place.}

An intriguing question is whether the \rzzz{inbetweening and lifting problems above} can be ``fused" to produce 4D contents, i.e., 3D object with motion, from a {\em minimalistic\/} and easy-to-acquire input setup. 
\rzzz{Indeed, with rapid advances in video foundation models and a push for {\em controllable\/} motion generation, there has been an emergence of works on video frame interpolation~\cite{reda2022film,feng2024explorative,danier2024ldmvfi,xing2024dynamicrafter,wang2025framer} which produce {\em video\/}, rather than 4D, outputs from two images.}

\begin{figure}[!t]
\centering
    \includegraphics[width=\linewidth]{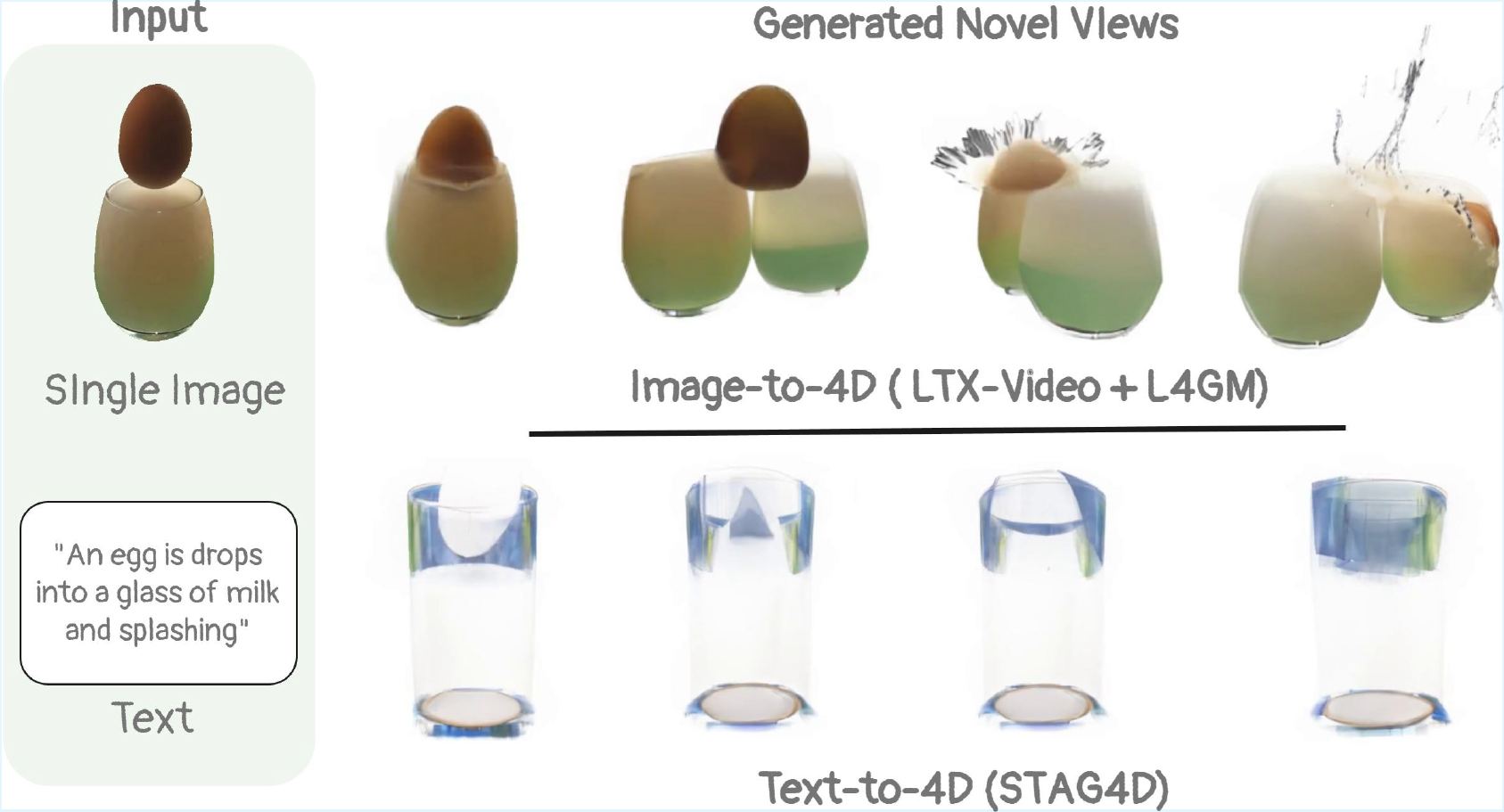}
    \caption{\rzzz{\textbf{Comparing to text-to-4D and single-image-to-4D.} Text-only and single-image inputs are insufficient to constrain the 4D generation by state-of-the-art models (L4GM \cite{ren2025l4gm,zeng2024stag4d} , leading to unnatural results and strong artifacts, especially for complex motions such as the one shown. Results from our method can be found in Fig.~\ref{fig:teaser}.}
    }
    \label{fig:comp_txt_single}
\end{figure}

In this paper, we seek a solution to the novel task of generating {\em 4D interpolative contents\/} from merely two {\em single-view images\/} capturing an object in two distinct motion states.
\rzzz{Compared to video or 4D generation from text only~\cite{singer2023text4d,bahmani20244d} or from a single image~\cite{ren2023dreamgaussian4d}, our interpolative task
anchored on two images can leverage more precise motion control to better constrain the generative process and lead to more plausible and cleaner results, as shown in Figs.~\ref{fig:teaser} and~\ref{fig:comp_txt_single}.}


We call our task and the proposed method both as {\em In-2-4D\/}, for Inbetweening from two (2) single-view images to 4D generation.
%
%
%
Aside from the sparse inputs, we aim to tackle additional challenges related to the {\em diversity\/} and {\em complexity\/} of the generated motions: 
a) no \am{particular} assumptions are made on the object or motion categories;
b) arbitrary {\em freeform\/} motions are permissable, without assumptions on rigidity or volume/topology preservation, e.g., \rzzz{see Fig.~\ref{fig:teaser} for a floral motion that is non-rigid and quite intricate, and 
an egg dropping into liquid, causing a splash and a topology change;} 
\rzz{c) moderately complex and long-range} motions are allowed, where the two input motion states are not assumed to be close in time, \rzzz{pose, or structure.}
Our goal is to synthesize a smooth and believable 3D transition between them.

\begin{figure*}[!t]
\centering
    \includegraphics[width=\textwidth]{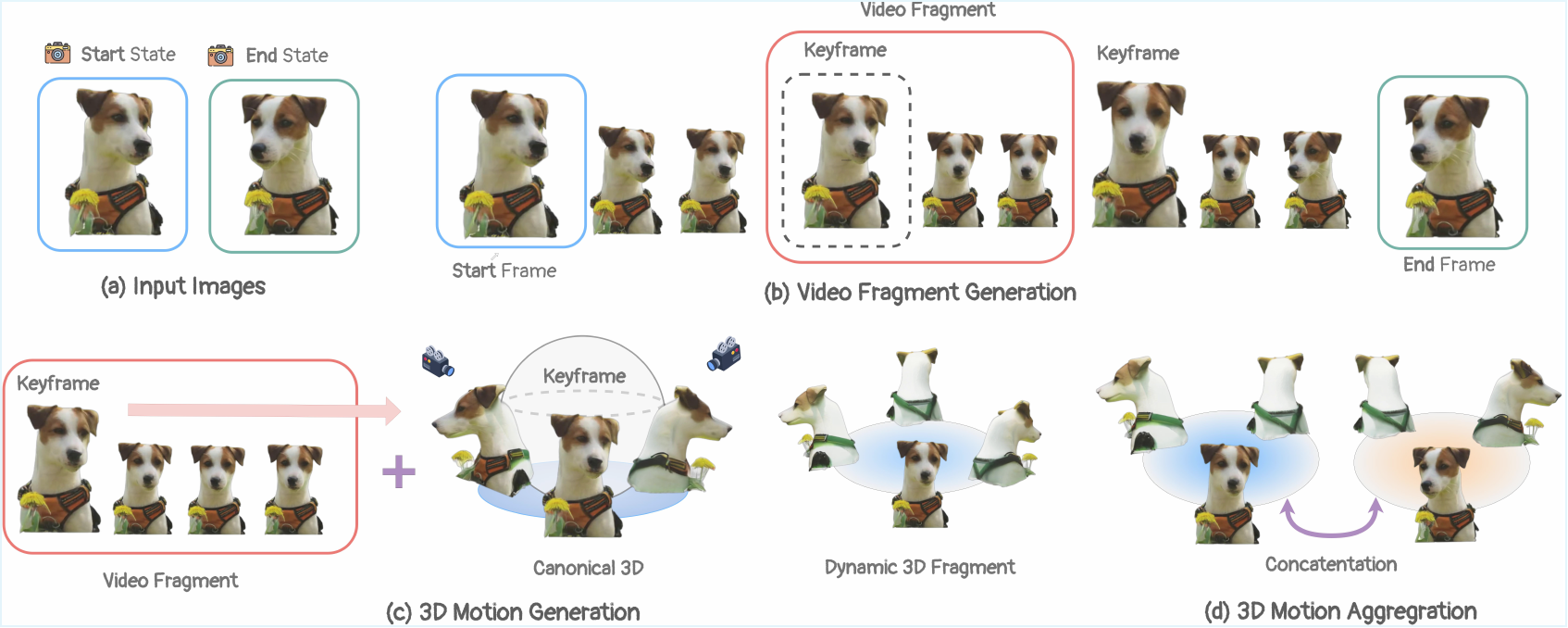}
    \caption{\textbf{Illustration of our \method{} solution pipeline.} 
    \am{Given two single-view images as input (a), we first generate {\em keyframes\/} \rzzz{inbetween them to bridge the motion gaps so as to avoid abrupt motions between consecutive keyframes to generate video fragments} (b)  . These keyframes are then utilized to learn static 3D geometries per fragment which are then deformed using a deformation field (e.g., Hexplane) to obtain a 4D scene \rzzz{in the form of 3D Gaussian splats (3DGS)} per fragment. Finally, we obtain a 4D scene \rzzz{in the form of 3D Gaussian splats (3DGS)} per fragment. (c)
 }        
    To aggregate the deformations, we linearly interpolate the deformation fields (d) in a cascading fashion and apply smoothing constraints on the \rzzz{3DGS to improve geometric and structural consistency between the generated novel views.}}
    \label{fig:pipeline}
\end{figure*}

At the high level, our method operates in two phases: 
2D still images to video via interpolation, and then video-to-4D via lifting, as illustrated in Fig.~\ref{fig:pipeline}.
To handle arbitrary and diverse motions, we leverage video foundational models. However, 
most such models are built on video diffusion~\cite{blattmann2023align}, which has been trained predominantly by short videos. As such, 
\rz{they can be ineffective for motion inbetweening when the input states span large geometry or structural changes, resulting in large motion ``jumps'' and absence of
detailed and intricate object movements.} \snn{When such video models are used to model complex motion, they often suffer from visible artifacts which further detoriate the novel views when lifted to 4D space.}

To this end, we develop a {\em divide-and-conquer\/} approach to adaptively and recursively generate a set of \snn{keyframes to form a coarse outline of the motion, breaking the long-range and nonlinear motion into simpler, temporally-local transitions that are easier to model and interpolate.}
To start, we employ a foundational video interpolation model such as 
\cready{DynamiCrafter \cite{xing2023dynamicrafter}} to generate an initial set of intermediate frames between the two input states. \rz{Then we
perform motion and appearance feature analyses over these frames to select one or more {\em keyframes\/} \am{that are visually close to the input
states and show significant motion jumps}. 
Consecutive keyframes that incur a large motion will anchor a new video interpolation to generate more immediate frames, effectively ``magnifying'' the
said motion. 
\am{This} process is carried out hierarchically until all motions between consecutive keyframes are sufficiently small. }
%
%
Notably, our \snn{video fragment generation is training-free and requires no fine-tuning of video diffusion models, making it both accessible and effective for 4D generation. For each video fragment, and in parallel with other fragments, we first learn a distinct {\em static\/} 3D Gaussian splatting (3DGS) model to capture the object geometry. We then apply a deformation field to convert this 3DGS into a dynamic, i.e., 4D, model by utilizing multi-view diffusion priors to refine the warping, geometry, and textures over unseen areas. By construction, the fragment contains relatively simple motions, hence multi-view generation can effectively mitigate texture and geometry degradation.}

Finally, we merge the independently generated 4D fragments  where we \am{first} linearly interpolate \am{and then optimize} the deformation fields over an overlapping frame and regularize the geometry of novel views 
in a cascading sliding window fashion to smooth the orientation of the dynamic 3DGS based on the neighboring frames. Fig.~\ref{fig:pipeline} overviews our pipeline. 

Our main contributions are summarized below:

\begin{itemize}
\item To the best of our knowledge, \method{} is the first method for generative 4D inbetweening over two distant monocular frames spanning arbitrary motions.
\item Our novel 
\am{hierarchical} approach breaks the complex inbetweening into a series of simpler motion estimations via \am{video}, and then 4D (i.e., dynamic 3DGS) \am{generation}. 
\item To generate smooth 3D object and motion transitions, we further optimize the 3D trajectories using a bottom-up merging strategy with smoothing regularization.
\item We contribute a new 4D interpolation benchmark \textbf{I4D-15} on challenging real world object motions. 
\end{itemize}


\am{We conduct extensive experiments on I4D-15 and \cready{Consistent4D \cite{jiang2023consistent4d} benchmark} for evaluation. Quantitative and qualitative comparisons are made to methods and baselines to demonstrate the effectiveness of our method in terms of the quality of generated results, generalizability, and handling of a variety of motions; see Fig.~\ref{fig:teaser}.} While achieving superior generation quality \rzzz{and remaining faithful to the input images than all current alternatives, our solution is far from artifact-free. This highlights the significant challenges that still lie ahead with 4D content generation from highly sparse inputs which still aim to exert intricate control.} As a first attempt at tackling such a complex problem, we hope it can serve as a promising start to stimulate future work.


%% file: main/02_lit.tex
\section{Related Work}
\label{sec:literature}

\rzzz{In computer graphics, most works on motion inbetweening have been developed for character animation, e.g.,~\cite{zhou2020generativetweeninglongterminbetweening,cohan2024MIBDiff,qin2022MIBTwoStage,zhong2025Sketch2Anim}.
Our work aims to handle arbitrary and diverse motions, leveraging foundational models, for the novel task of In-2-4D. In this section, we mainly cover related works on 4D generation and video/4D interpolation.
}

\vspace{2pt}

\noindent\textbf{4D dynamic scene generation.}
Recent works \cite{wu20234dgaussians, yang2023deformable3dgs, Liang2023GauFReGD, Guo2024Motionaware3G, zeng2024stag4d, ren2023dreamgaussian4d, wu2024sc4d} extend 3D Gaussian Splatting (3DGS)~\cite{kerbl20233Dgaussians} to 4D using time-conditioned deformation networks with \rzzz{Score Distillation Sampling (SDS)} and multi-view geometry.  MAV3D \cite{singer2023text4d} pioneered text-to-4D via NeRF and score distillation, followed by similar approaches~\cite{bahmani20244d}. Consistent4D~\cite{jiang2023consistent4d} introduces video-to-4D with pre-trained image diffusion models, extending to video-conditional 4D generation~\cite{shi2023zero123++, wang2023imagedream, yang2024diffusion}. STAG4D~\cite{zeng2024stag4d} and 4DGen~\cite{yin20234dgen} refine diffusion with pseudo-labels, while SC4D~\cite{wu2024sc4d} employs sparse Gaussians and \rzzz{Linear Binding Skinning (LBS)} for dynamic 3D \rzzz{generation}. 
\cready{Recently, research has shifted towards SDS-free 4D foundation models \cite{ren2025l4gm,xie2024sv4d,wu2025cat4d,zhang20244diffusion,jiang2024animate3d,liang2024diffusion4d,sun2024eg4d} that synthesize synchronized multi-view videos as intermediates for 4D reconstruction, bypassing SDS optimization. While these models produce plausible 4D motions, they remain limited by the inductive bias of their training data and fail to generalize to scenes with complex dynamics. Despite progress in video diffusion, current video-to-4D methods still struggle with highly dynamic motions and accumulate errors over long sequences due to reliance on a single canonical model. To address this, our \rzzz{framework} segments videos into shorter fragments, \rzzz{via keyframe generation}, each associated with its own 3D canonical model, thereby improving geometry.}


\vspace{1.5pt}

\noindent\textbf{Video inbetweening \rzzz{or frame interpolation.}}
\rzzz{Motion inbetweening allows finer-grained control compared to textual inputs or motion extrapolation from a single frame.}
\am{Recent methods have extended pre-trained diffusion-based text-to-image models to generate motion from static images by adapting UNets to temporal data~\cite{wu2023tuneavideo,text2video-zero,singer2022makeavideo}. One notable model is AnimateDiff~\cite{guo2023animatediff}, which learns low-rank adapters for diverse motion patterns. More recent approaches condition pre-trained text-to-video models on input images. VideoCrafter1~\cite{chen2023videocrafter1} uses dual cross-attention layers to combine image features with text prompts, while DynamiCrafter~\cite{xing2023dynamicrafter} further refines this by concatenating the input image with noisy latent features.
Our method builds on DynamiCrafter to enhance its outputs through recursive video magnification. While several video magnification techniques exist~\cite{video-mag-survey,Motion-mag-durand}, we leverage a video inbetweening network (e.g., DynamiCrafter) to interpolate frames and amplify motion when large displacements are present. Decomposing large motions into smaller fragments with smoother transitions reduces geometric ambiguities between consecutive frames, producing 4D results with fewer artifacts and improved visual quality.}

\vspace{2pt}

\noindent\textbf{4D scene interpolation.} 
Lately, dynamic 3D scene interpolation, or 4D Interpolation, has become more popular. Earlier works~\cite{park2023temporal} leverage neural radiance fields (NeRF) for temporally coherent 3D reconstructions, while NeuralPCI~\cite{zheng2023neuralpci} employs neural fields for multi-frame, non-linear 3D point cloud interpolation. PAPR~\cite{peng2024papr} estimates motion via point-based rendering and local displacement optimization. Recent methods~\cite{ling2024align,jiang2023consistent4d,ren2025l4gm} use frame motions from Diffusion-based Video Interpolation models~\cite{blattmann2023stable,xing2023dynamicrafter} to infer 3D deformation. However, Video Diffusion models~\cite{blattmann2023align}, trained on short clips (e.g., 16 frames), struggle with long sequences, causing artifacts from large per-frame motion jumps. To address this, we use generative Video Interpolation models (e.g., DynamiCrafter~\cite{xing2023dynamicrafter}) hierarchically for longer 3D trajectory estimation without extra training.
\rzzz{In addition, In-2-4D represents the first attempt at general 4D interpolation from two single-view motion frames.}

%% file: main/03_model.tex
\section{Methodology}
\label{sec:method}


\am{An overview of our method is shown in Fig.~\ref{fig:pipeline}. Given two images representing the start and end states of an object in motion, we aim to generate and reconstruct the motion in 4D (3D+motion). To predict the motion, we use a video interpolation model, but large motions between frames can lead to ambiguous interpretations and results with artifacts. To solve this, we employ a hierarchical approach to identify keyframes that are visually close to the input states and exhibit significant motion, then generate smooth fragments between them. For each fragment, the 3D representation of the keyframe is first constructed using Gaussian Splatting. The temporal frames within the fragment serve as motion guidance, enabling their transformation into dynamic Gaussians through a deformation field. To enhance temporal consistency and refine 3D motion of the fragment, we expand the self-attention of multi-view diffusion across time steps and introduce rigid transformation regularization. Finally, the independently generated 3D motion segments are merged by interpolating the boundary deformation fields and optimizing them to align with the guiding video. This ensures flicker-free transitions.}

\subsection{Problem Setup}
\noindent \textbf{Task description.} 
\am{Given a pair of start and end single view images $I_{s}$ and $I_{e}$ $\in \mathbb{R}^{H \times W \times 3}$ representing a dynamic scene possibly having a complex and large motion, our task is to generate a 4D interpolated scene that can be observed at any point of time or view.}


\noindent \textbf{Our framework.}
\saura{
Our objective is to generate smooth motion while minimizing 3D artifacts in novel views. To achieve this, we introduce gradual local displacements and insert frames in regions with complex motion to prevent abrupt transitions. First, keyframes are adaptively generated by analyzing motion differences in feature space, segmenting fragments with simple motion (Sec.\ref{sec:Hier_Subse_Gen}). These fragments are then individually lifted to 4D space using their respective motion (Sec.\ref{sec:Multi-Can_Traj_Estimation}). Finally, local deformations are integrated into a globally smooth 4D motion with regularization (Sec.~\ref{sec:cas_3D_mot_Agg}).
}


\begin{figure}[t]
\centering
\includegraphics[width=\linewidth]{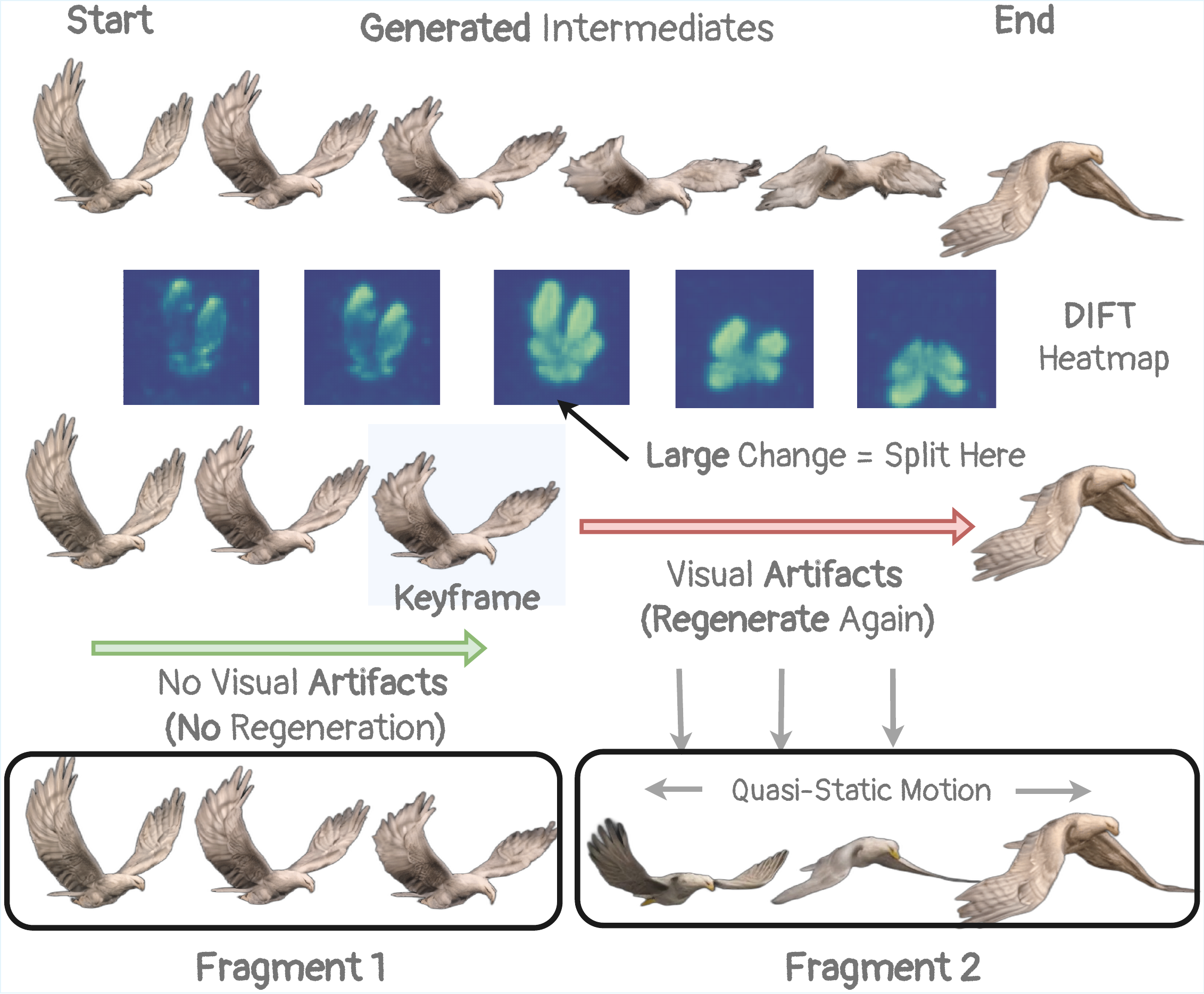}
\caption{\textbf{Illustration of Hierarchal Fragment Generation.} \saura{At each generation step, a keyframe is selected by finding the largest motion from the DIFT heatmap and FID score. New frames are re-generated using the keyframe to minimize large motion changes. Selection of the keyframes are done in a hierarchial manner to generate fragments having simple motions}} 
    \label{fig:hierarchy}
\end{figure}

\subsection{Temporal Fragment Hierarchy} 
\label{sec:Hier_Subse_Gen}
\begin{figure}[t]
\centering
    \includegraphics[width=\linewidth]{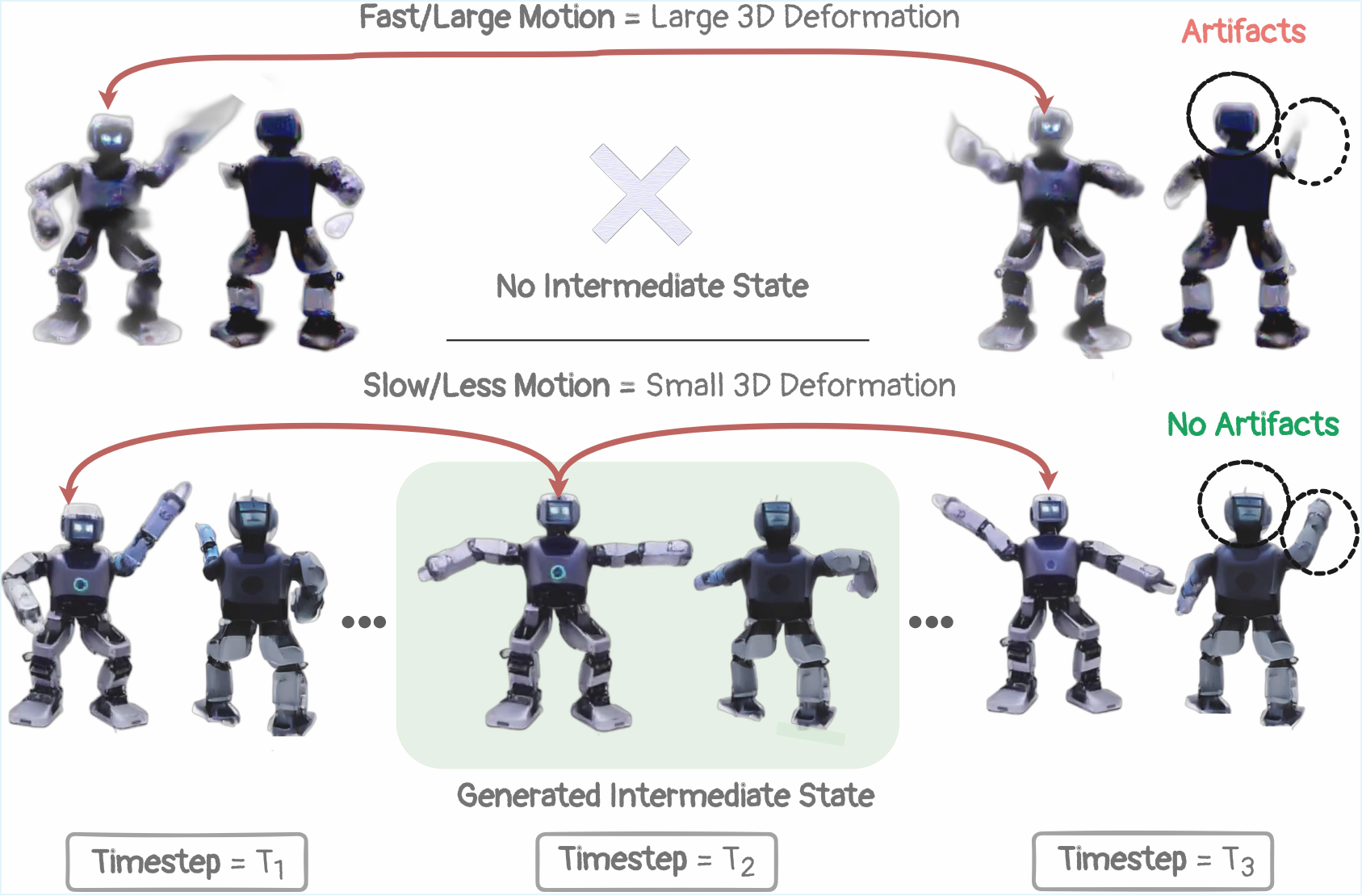}
    \caption{\textbf{Effect of inbetweening on geometry.}
    \saura{\am{When the input states are significantly different}, the 3D deformation module undergoes large movements (fast motion) leading to artifacts in novel views, whereas generating intermediate frames between the states (slow motion) enhances the geometry using smaller deformations.}}
    \label{fig:slow_geom}
\end{figure}

\saura{We propose a method for identifying keyframes in fragments with significant deformations and adaptively expanding them. Large deformations between start and end states induce rapid intermediate motion changes, which hinder 3D deformation learning \cite{liang2024monocular} as shown in Fig.~\ref{fig:slow_geom}. To mitigate this, we partition the motion trajectory into fragments with smoother \saura{quasi-static} motions, selecting keyframes densely in dynamic regions and sparsely in static regions. This balances training overhead, model size, and performance and enhances temporal consistency.}

\noindent\textbf{Hierarchical key-frame generator.}
\saura{To generate keyframes for \am{the} intermediate motion between two initial states, we employ a Video Interpolation Model (e.g., DynamiCrafter), denoted as $\psi(.)$. Given input images $I_s$ and $I_e$ along with a motion prompt $p$ (extracted using BLIP \cite{li2022blip}), we generate a sequence of latent frames $Z$. The pairwise DIFT \cite{tang2023emergent} features quantify frame-wise similarity, enabling a rapid assessment of motion changes. As illustrated in Fig.~\ref{fig:hierarchy}, a heatmap visualizes temporal variations, where significant object movements or new appearances \am{are represented} as bright regions.
The heatmap between frames $I_i$ and $I_j$ is computed as:
\begin{equation*} \begin{aligned} H^{p}_{i,j} = \texttt{CS}(f^{p}_{i},f^{q*}_{j}), \text{where} \ q^{*} = \arg\max \texttt{CS}(f^{p}_{i},f^{q}_{j}) \end{aligned} \end{equation*}
where $\texttt{CS}(.)$ represents cosine similarity, and $p, q$ denote tokens of DIFT feature $f$. A frame is marked as a keyframe if the mean heatmap value of $H_{i,j}$ between frame pairs $I_{i},I_{j}$ falls below a predefined threshold.
To sample the best keyframe in terms of visual fidelity we further assess its consistency with the initial inputs using FID metric. 
The keyframe latent $z_m$ at timestep $m$ is selected based on the highest FID against the input states to remain faithful to inputs. For instance, in \cready{Fig~\ref{fig:hierarchy}}, the chosen keyframe exhibits the highest fidelity to the input states of the eagle.
Once identified, the keyframe divides the motion trajectory into two segments: ${z_{s},z_{m}}$ and ${z_{m},z_{e}}$. The interpolation model $\psi(.)$ then utilizes these fragments iteratively in a "divide-and-conquer" fashion, identifying further keyframes until the full video is processed. This hierarchical approach ensures adaptive keyframe density, reducing redundant intermediate frames in low-motion areas while preserving complex motion details.
\am{Therefore, the hierarchical keyframe selection is performed recursively based on prior selections:}
\begin{equation} 
\mathcal{K} = {K_{(s)(1)}, K_{(1)(2)}, K_{(2)(3)}, ..., K_{(c)(e)}} 
\end{equation}
where $K_{(i)(j)}$ denotes the keyframe between states $i$ and $j$.}

\noindent\textbf{Temporal fragment generation.} \am{Having keyframes $\mathcal{K}$, we reuse the video interpolation module $\psi(.)$ to perform inebetweening for consecutive keyframes $K_{(i)(j)}$ and $K_{(j)(k)}$. \am{Since $\psi(.)$ receives latents, we interpolate the latents and decode them using a VAE decoder to insert new RGB frames.} Since the consecutive keyframes represent simple quasi-static motions, this interpolation generates smooth fragments with \am{fewer} artifacts. We generate $T$ such fragments denoted by $\mathcal{V}_{i}$ each having fixed number of frames $f$ (e.g., 16) representing the motion between keyframes:} 
\begin{equation}
    \mathbf{V} = \{ \mathcal{V}_{s(1f)}, \mathcal{V}_{(1f)(2f)}, ... ,\mathcal{V}_{((c-1)f)e} \},
\label{eq:V}
\end{equation} 
\am{where $\mathcal{V}_{s(1f)} = \mathcal{D}(\psi(z_{(s)(1)},z_{(1)(2)}))$; $\mathcal{D}$ is VAE  decoder.}


\subsection{Modelling Intra-Fragment Geometry}
\label{sec:Multi-Can_Traj_Estimation}
\saura{We lift individual video fragments to 4D by generating multi-view videos of the object. Existing video-to-4D methods~\cite{ren2023dreamgaussian4d,wu2024sc4d} \am{use} multi-view diffusion models~\cite{liu2023zero} to synthesize multi-view videos by independently processing each frame. However, this approach ensures cross-view consistency but leads to temporally inconsistent geometry Fig~\ref{fig:intra_frag}. Moreover, due to sole reliance on multi-view video supervision, Gaussian splatting often produces flickering and texture variations \cite{luiten2023dynamic} due to its high degrees of freedom per point and lack of motion constraints. We address these issues by generating temporally consistent multi-view videos and regularizing motion with rigid constraints within each fragment.}

\noindent\textbf{Learning canonical 3D.} 
We start by estimating a canonical Gaussian representation and then add motion to it from the multi-view videos.
\saura{For each temporal fragment $\mathcal{V}_{i}$, we designate the keyframe $K_{(i)(j)}$ as the canonical reference and reconstruct its 3D structure via multi-view synthesis. Specifically, we employ the multiview diffusion model Era3D \cite{li2024era3d} to generate multi-view images from $K_{(i)(j)}$, followed by 3DGS \cite{kerbl20233Dgaussians} for coarse geometry reconstruction. As each fragment is processed independently, parallel execution reduces computation time. The resulting coarse geometry provides an effective initialization for learning texture and geometry across the remaining temporal frames.}

\noindent\textbf{Dynamic 3D fragment generation.} 
\saura{After learning 3D static Gaussians, we leverage motion priors from the video fragment to transform them into dynamic Gaussians. Since single-view videos cannot provide diverse observations of the scene from different viewpoints, we use multi-view videos. To promote temporal consistency, rather than generating multiple-view of the frames independently at each timestep, we propagate the self-attention features of the multi-view diffusion model \cite{li2024era3d} from the canonical frame across the entire frames of the fragment as follows:}
\begin{equation*}
\begin{split}
    &{z}_{t} \leftarrow \gamma . {z}_{c} + (1 - \gamma).{z}_{t}, \\
    &\mathcal{Q} = \mathcal{W}^{q}.{z}_{t}, \mathcal{K} = \mathcal{W}^{k}.{z}_{t}, \mathcal{V}=\mathcal{W}^{v}.{z}_{t}, \\
    &\texttt{Attention}(\mathcal{Q},\mathcal{K},\mathcal{V}) = Softmax(\frac{\mathcal{Q}\mathcal{K}^{T}}{\sqrt{d_{k}}}.\mathcal{V}).
\end{split}
\end{equation*}
\saura{where ${z}_{c}$ is the multi-view latent of the canonical frame, $t$ is timestep of the video fragment, $\gamma$ is the blending weight and $d_{k}$ is the key dimension. 
 With quasi-static motions in each video fragment, the generated multi-view videos have minimal variation in viewpoints, making it easier for the model to capture accurate and consistent geometry (Fig.~\ref{fig:intra_frag}). With the synthesized multi-view videos of the dynamic
object, we optimize a 3D deformation field (denoted by $\Delta_{\boldsymbol{\Phi}_i}$)  to enable free-viewpoint rendering. We chose Hexplanes \cite{Cao2023HEXPLANE} as our deformation field due to modeling efficiency. The deformation field predicts each Gaussian’s geometric offsets at a given timestamp relative to the mean canonical state (keyframe). For each timestamp $\tau$ of video and 3D Gaussian $p$, Hexplanes predict displacement, rotation, and scaling for the 3D gaussian points.}

\noindent \textbf{Optimization objective.} \am{To respect the driving video and optimize the deformation field, we fix the camera to a view and minimize the Mean Squared Error (MSE) between the rendered image and each video frame:
\begin{equation}
    \mathcal{L}_\textrm{Ref} = \frac{1}{\mathcal{T}}\sum_{\tau=1}^{\mathcal{T}}||f(\phi(S, \tau),o_\textrm{Ref}) - I^\tau_\textrm{Ref}||^2_2,
\end{equation}
where $I^\tau_\textrm{Ref}$ is the $\tau$-th frame, $o_\textrm{Ref}$ is the reference viewpoint, and $f$ is the rendering function. 
\saura{Dynamic Gaussians tend to move freely across regions of similar color~\cite{luiten2023dynamic} without constraints, causing flickering and floating artifacts that degrade 4D motion realism. As motion within each fragment is minimal, we enforce rigid assumptions on point movements relative to the canonical state \am{by regularization:}
\begin{equation}
\mathcal{L}_{rigid} = || d(\mu^{c}_{i} , \mu^{c}_{j}) - d(\mu^{\tau}_{i},\mu^{\tau}_{j})||_{1}
\end{equation}
where $d(x , y) = ||x - y||_{2}$ is the distance function, and $\mu$ denotes the Gaussian center of neighboring clusters $\mathcal{N}$. $\mu^{c}$ and $\mu^{\tau}$ represent Gaussian centers in the canonical and arbitrary timestep frames, respectively. This regularization permits non-rigid deformations (like bending) while minimizing local rigid distortions.}
}
\am{In addition to this, we also use a random view at each timestep and apply foreground mask loss $\mathcal{L}_{mask}$, resulting in a total training objective:
\begin{equation}
\label{eq:L}
    \mathcal{L} = \lambda_{1}\mathcal{L}_{Ref} + \lambda_{2}\mathcal{L}_{mask} + \lambda_{3}\mathcal{L}_{rigid},
\end{equation}
where \cready{$\lambda_{1}$, $\lambda_{2}$ and $\lambda_{3}$ are loss weights which are set as $0.4, 0.3, 0.3$ respectively}. In training, for each fragment $\mathcal{V}_{i}$, we first use $\mathcal{L}$ to supervise the static 3D Gaussian, then train the dynamic 4D Gaussian with all reference frames.}

\subsection{Cascaded 3D Motion Aggregation}
\label{sec:cas_3D_mot_Agg}
\am{Since we learn each fragment’s deformation independently, the entire video may lack consistency over the global geometry and motion. The overall 3D deformation field $\Delta$ consists of mini-deformations per fragment: 
\begin{equation}
    \Delta = [\boldsymbol{\Delta_{\boldsymbol{\Phi}_1}},\boldsymbol{\Delta_{\boldsymbol{\Phi}_2}},...,\boldsymbol{\Delta_{\boldsymbol{\Phi}_K}}],
\end{equation}
where each $\boldsymbol{\Delta_{\boldsymbol{\Phi}_i}}$ is optimized separately. To achieve smooth, flicker-free motion in novel views, we need to merge these fragment deformations.}


\begin{figure}[t]
\centering
    \includegraphics[width=\linewidth]{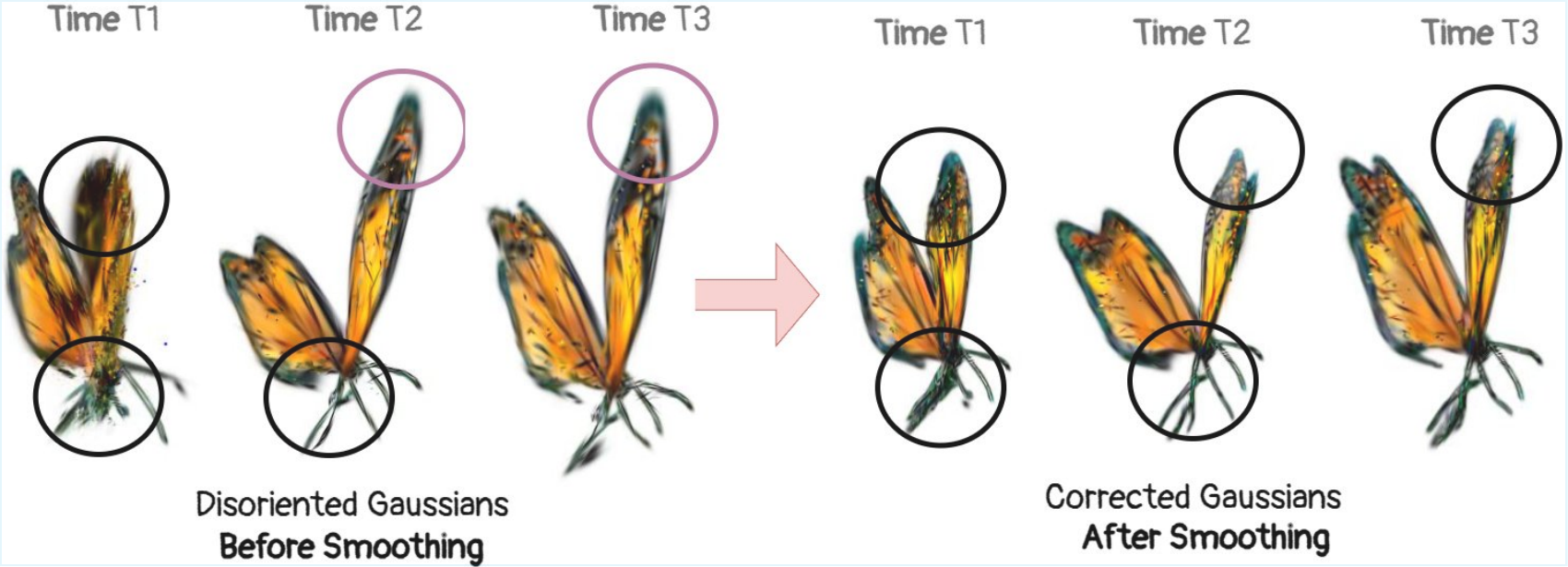}
    \caption{\textbf{Trajectory smoothing} of fragments leads to correction of Gaussians and help render better novel views.}
    \label{fig:merge}
\end{figure}

\noindent\textbf{Motion merging.}
\am{With overlapping frames between adjacent fragments, we linearly interpolate the deformation fields for these overlaps. Specifically, we define the interpolated deformation field as: 
\begin{equation} \boldsymbol{\Delta}^{\textrm{merge}}{\boldsymbol{\Phi}_{ij}} = \lambda\boldsymbol{\Delta}{\boldsymbol{\Phi}_i} + (1- \lambda)\boldsymbol{\Delta^{*}}{\boldsymbol{\Phi}_j}, \end{equation} 
where $\lambda = 0.5$, and only $\boldsymbol{\Delta^{*}}{\boldsymbol{\Phi}_j}$ is learnable. 
With intra-frame motions already smooth, we freeze $\boldsymbol{\Delta}{\boldsymbol{\Phi}_{i}}$ and $\boldsymbol{\Delta}{\boldsymbol{\Phi}_{j}}$, optimizing only $\boldsymbol{\Delta}^{\textrm{merge}}{\boldsymbol{\Phi}_{ij}}$ for a few iterations (e.g., 1,000) at a low learning rate using Eq.\ref{eq:L} to ensure smooth inter-frame motion between fragments.}
\am{Starting with $\boldsymbol{\Delta}{\boldsymbol{\Phi}_{1}}$, we progressively merge all deformation fields in a \emph{cumulative} fashion, resulting in a smooth and globally coherent 3D motionwithout abrupt transitions or flickers.}

\noindent\textbf{Cascaded trajectory smoothing.} \am{Despite smooth transitions across fragments, minor 3D inconsistencies may persist (see \ref{fig:merge}), often due to disoriented Gaussians causing blurry or over-reconstructed artifacts~\cite{kerbl20233d}. Since the 3D Gaussian geometry is controlled by the covariance matrix (i.e., rotation $q$ and scaling $s$), we regularize $q$ and $s$ over a fixed window with neighboring frame constraints. We adopt off-the-shelf tracking (e.g., CoTracker~\cite{karaev2023cotracker}) and depth models (e.g., DepthAnything~\cite{yang2024depth}) in a sliding window manner for post-processing. Given a window $w$, we estimate depth $\mathcal{D}$ and trajectory $\mathcal{T}$ on the video $\mathbf{V}$ (Eq.~\ref{eq:V}) and lift $N$ randomly selected trajectories to 3D using camera intrinsics. Points visible for at least $80\%$ frames are smoothed with Exponential Moving Average (EMA) on rotation and scale:
\begin{equation}
    \begin{split}
        q_{t} &= \frac{sin(\alpha.\theta)}{sin \theta} q_{t} + \frac{sin((1- \alpha).\theta)}{sin \theta} q_{t-1}, \\
        s_{t} &= \alpha s_{t} + (1- \alpha) s_{t-1},
    \end{split}
\end{equation}
where $\theta$ is the angle between consecutive rotations, and $\alpha$ is the EMA decay factor \cready{(set as 0.6)}. The process is repeated
iteratively until all disoriented Gaussians are corrected, yielding stable and flicker-free 3D reconstructions.}

%% file: main/04_exp.tex
\section{Experiments and Results}
\label{sec:experiment}
\noindent \textbf{Dataset.} 
\am{We evaluate our method on 4D sequences with large object motions, defined when multiple object parts move. \saura{We introduce the \textit{I4D-15} benchmark, comprising 15 articulated objects across categories like Vehicles, Flowers, Humans, Animals, and Daily Life scenes. The dataset includes 64-frame sequences at 16 fps from Objaverse1.0~\cite{deitke2023objaverse} and SketchFab \cite{spiess2024sketchfab}, rendered from 5 evenly spaced views at $0^{\circ}$ elevation. We select the first and last frames of the front view as input states and reserve the remaining video for evaluation. The camera radius is $1.5$, and the FOV is $49.1^{\circ}$, consistent with~\cite{jiang2023consistent4d}. Motion filtering~\cite{li2024puppet} ensures large motion selection. Evaluation uses appearance metrics (LPIPS, FVD)\cite{ren2023dreamgaussian4d} and geometry metrics (SI-CD, CD)\cite{peng2024papr}.}\cready{In addition to this, we have also evaluated our method along with the baselines on Consistent-4D \cite{jiang2023consistent4d} dataset for more extensive validation.}}

\begin{figure}[t]
\centering
    \includegraphics[width=\linewidth]{img/ablation_v7.pdf}
    \caption{\textbf{Effect of Inter-Fragment Consistency.}
    \saura{Without using any consistency or regularization, blurriness and oversaturated artifacts are produced. Rigid consistency improves the structure and when combined with temporal-aware multi-view generation, better geometry and texture are obtained.}}
    \label{fig:intra_frag}
\end{figure}

\begin{figure*}[t]
    \includegraphics[width=\linewidth]{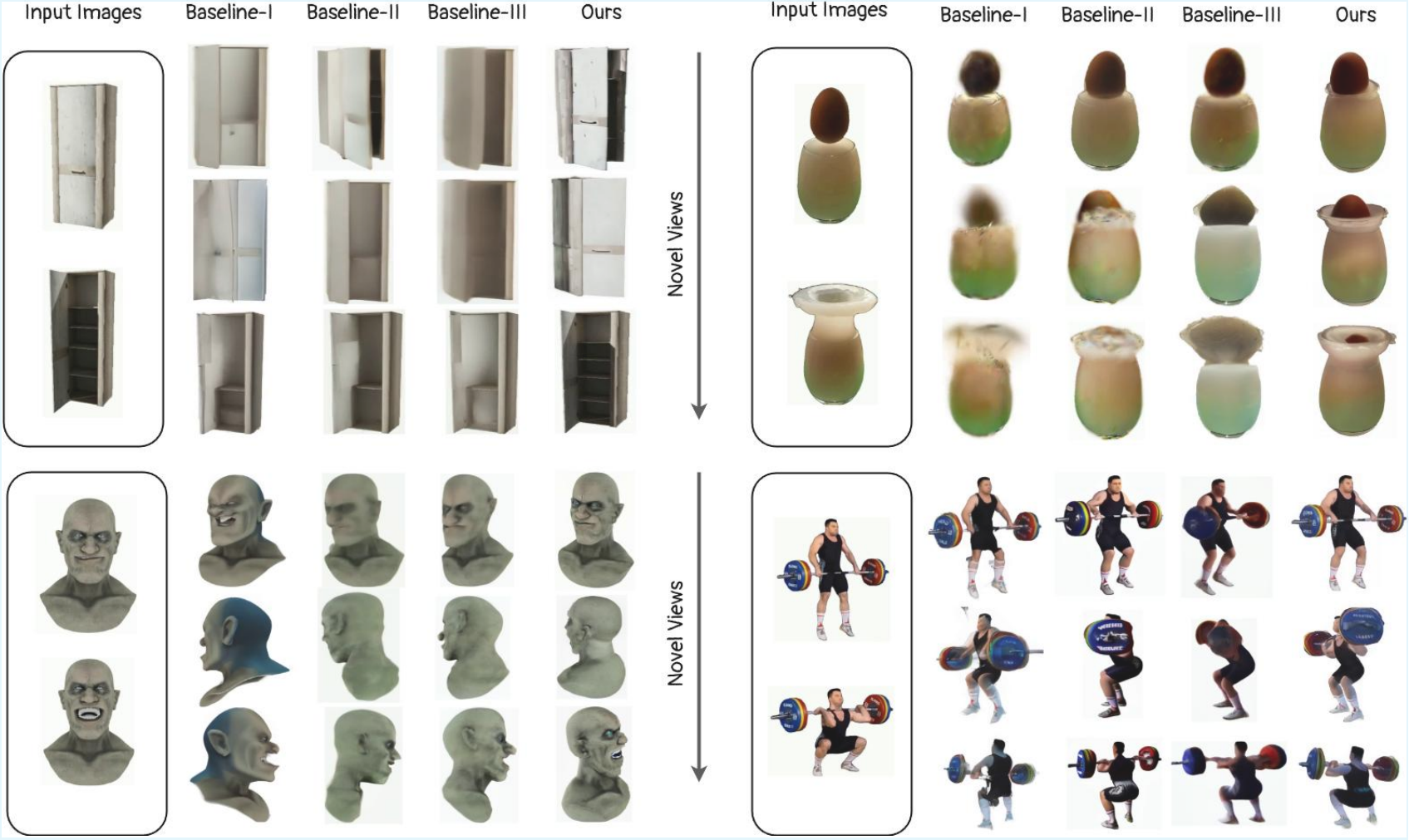}
    \caption{\cready{Qualitative comparison of our approach with the baselines on challenging synthetic and real-world scenes (e.g, water splash , weightlifting).}}
    \label{fig:main_qual}
\end{figure*}
\noindent \textbf{Baselines.} \am{As we introduce a novel task, no existing method can be directly applied to our setting. Thus, we establish the following baselines for quantitative comparison: For 4D baseline generation, we 
first generate videos  from two frames and subsequently lift it to 4D using a Video-to-4D approach. (a) \textit{Baseline-I} employs a image based interpolation \cready{DreamMover \cite{shen2024dreammover} to generate keyframes} followed by \cite{xing2024dynamicrafter} to generate video fragments and then use a latest Video-to-4D method \cite{ren2025l4gm} (b) \textit{Baseline-II} utilizes latest video interpolation approach \cite{xing2024dynamicrafter} along with our temporal \cready{fragment generation} module for 2D interpolation and a recent \cready{Video-to-4D method \cite{xie2024sv4d}. (c) \textit{Baseline-III} is an improved version of Baseline I in two ways: first, it incorporates the output of our Temporal Fragment Hierarchy Module (Sec. 3.2) for video generation and second, we replace L4GM \cite{ren2025l4gm} with a more recent video-to-4D model, SV4D \cite{xie2024sv4d} to lift the video to 4D.} }
\saura{Additionally, we evaluate a single-image-to-4D task on our dataset, with further analysis provided in the supplementary.} Baseline results are obtained using official GitHub implementations.

\noindent\textbf{Quantitative comparisons.} \am{
\saura{We quantitatively evaluate our approach on our I4D-15 benchmark. Two images from one view are used as input and 4 videos (each 64 frames) from other viewpoints and their corresponding timsetep point clouds are used for evaluation.}
As shown in Tab.~\ref{tab:IfD}, \saura{our method outperforms the baseline across all metrics in appearance and geometry.} \cready{To compare our method on the Consistent4D benchmark, we use the first and the last frame of the input test video and use the novel view videos of the test set for evaluation. Similar observations can be drawn from Tab.~\ref{tab:C4D}, where our method is much superior than the baselines which is expected as the dataset contains simple motions.} This shows the robustness of our model design and effectiveness of our method in \saura{handling complex motions in 4D by dividing it into quasi-static temporal fragments}.}

\begin{table}[t]
\centering
\setlength{\tabcolsep}{5pt}
\small
\caption{Quantitative Analysis on proposed I4D-15 Dataset.}
\vspace{-0.15in}
\label{tab:IfD}
\cready{
\begin{tabular}{c|ccc|cc}
\hline
                         & \multicolumn{3}{c|}{Appearance} & \multicolumn{2}{c}{Geometry} \\ \cline{2-6} 
\multirow{-2}{*}{Method} & CLIP $\uparrow$      & LPIPS $\downarrow$     & FVD $\downarrow$    & SI-CD $\downarrow$         & CD  $\downarrow$        \\ \hline
Baseline-I              & 0.81        & 0.143        & 992.23      & 33.58             & 0.76          \\
Baseline-II             & 0.84        & 0.136        & 729.32      & 31.79             & 0.73          \\
Baseline-III            & 0.83        & 0.138        & 811.26      & 32.46             & 0.74          \\\hline
\rowcolor[HTML]{CBCEFB} 
\bf Ours                     & \bf 0.91        & \bf 0.103        & \bf 679.23      & \bf 22.67             & \bf 0.59          \\ \hline
\end{tabular}
\vspace{-0.1in}
}
\end{table}

\begin{table}[t]
\centering
\setlength{\tabcolsep}{10pt}
\small
\caption{Quantitative Analysis on proposed Consistent-4D Dataset.}
\vspace{-0.15in}
\label{tab:C4D}
\cready{
\begin{tabular}{c|ccc}
\hline
Method        & LPIPS $\downarrow$         & CLIP-S $\uparrow$       & FVD $\downarrow$             \\ \hline
Baseline-I    & 0.18          & 0.84          & 925.46          \\
Baseline-II   & 0.14          & 0.87          & 810.24          \\
Baseline-III  & 0.16          & 0.86          & 878.13          \\ \hline
\rowcolor[HTML]{CBCEFB} 
\textbf{Ours} & \textbf{0.13} & \textbf{0.90} & \textbf{741.88} \\ \hline
\end{tabular}
\vspace{-0.1in}
}
\end{table}

\noindent\textbf{Qualitative comparisons.} \am{Fig.~\ref{fig:main_qual} \cready{and Fig ~\ref{fig:qual-c4d}} provides some qualitative comparisons with the baseline \cready{on both real-world scenes and synthetic objects}. It is apparent that our method produces fewer artifacts \cready{in the novel views and also preserve the geometry better in complex motions}. 
\cready{In addition to this, we also present a gallery visualization of scenes in Fig~\ref{fig:qual-gallery} for different motions and object categories from our proposed I4D-15 benchmark and additional real-world scenes.} \cready{More videos are} provided in supplementary.
}


\noindent\textbf{Ablation Study.}
\am{This study evaluates the contribution of key components in our method on the I4D-15 benchmark. As shown in Tab~\ref{tab:vid-mag}, \saura{using four segments enables our model to decompose complex motion into finer details using our \cready{Temporal Fragment Hierarchy (TFH) module}, achieving the best cost-performance balance.} \saura{Additionally, we visually analyze the effect of Intra-Fragment consistency (Sec.~\ref{sec:Multi-Can_Traj_Estimation}) in Fig.~\ref{fig:intra_frag}, revealing that mv-consistency significantly enhances novel view synthesis, while rigid consistency mitigates deformation artifacts.} Tab.~\ref{tab:abl-moti} further highlights the impact of 3D motion aggregation. The combination of \saura{merging and smoothing} improves both appearance and geometry metrics, except for a slight decline in FVD when trajectory smoothing is applied. \saura{Moreover, we benchmark runtime against all baselines, as shown in Tab.~\ref{tab:runtime}, demonstrating that our approach outperforms the fastest baseline (B-I) by 25\% on an NVIDIA A100 GPU.} Additional ablations are provided in the supplementary material. }

\begin{table}[t]
\small
\setlength{\tabcolsep}{4pt}
\caption{\textbf{Ablation} on the impact of temporal fragment generation.}
\vspace{-0.15in}
\label{tab:vid-mag}
\centering
\begin{tabular}{c|cc|cc|c}
\hline
\# Fragments                 & LPIPS $\downarrow$         & FVD  $\downarrow$       & SI-CD $\downarrow$         & CD $\downarrow$ & Time $\downarrow$           \\ \hline
w/o TFH & 0.137          & 922.16          & 32.56          & 0.74 & 5 mins          \\ \hline
2  & 0.124          & 898.23          & 29.68          & 0.70  & 8 mins        \\
8  & \bf 0.101          & 680.11          & \bf 22.59          & 0.60 & 36 mins                  \\ \hline

\rowcolor[HTML]{CBCEFB} 
\textbf{4 (Ours)}          & 0.103 & \textbf{679.23} & 22.67 & \textbf{0.59} &\textbf{17 mins} \\ \hline
\end{tabular}
\end{table}

\begin{table}[t]
\small
\setlength{\tabcolsep}{5pt}
\caption{\textbf{Ablation} on the components of motion aggregation.}
\vspace{-0.15in}
\label{tab:abl-moti}
\centering
\begin{tabular}{cc|cc|cc}
\hline
\multirow{2}{*}{\begin{tabular}[c]{@{}c@{}}Motion \\ Merging\end{tabular}} & \multirow{2}{*}{\begin{tabular}[c]{@{}c@{}}Trajectory \\ Smoothing\end{tabular}} & \multicolumn{2}{c|}{Appearance} & \multicolumn{2}{c}{Geometry} \\ \cline{3-6} 
                                                                           &                                                                                  & LPIPS $\downarrow$           & FVD $\downarrow$           & SI-CD $\downarrow$          & CD $\downarrow$          \\ \hline
                                                                           \rowcolor[HTML]{CBCEFB} 
\cmark                                                                        & \cmark                                                                              & \textbf{0.103 }             & \textbf{679.23}            & \textbf{22.67}             & \textbf{0.59}          \\ \hline
\cmark                                                                        & \xmark                                                                               & 0.116              & 783.28            & 25.40             & 0.71          \\
\xmark                                                                         & \xmark                                                                                            & 0.137            & 922.16             & 32.56    & 0.74       \\ \hline
\end{tabular}
\end{table}

\noindent \textbf{User study.}
\saura{A user study was conducted, as human judgment is most effective for assessing 3D generation and motion quality. \cready{Since no standard metrics exist for evaluating 4D performance holistically, human judgment becomes a viable option. The study involved $20$ graduate students with a background in computer science, who were shown both the start/end state images and the output rendered videos from all the methods. Each object motion was presented from four canonical views (front, back, left, and right) and the participants were asked to rank four methods (1 = best, 4 = worst) based on how well the start/end state images align with the generated 3D geometry and how consistent is the motion (reduced flicker). In case of ties in motion consistency, 3D generation quality was prioritized. As shown in Tab.~\ref{tab:user}, our method achieves the best average rank (1.46), demonstrating the superiority of our model for overall 4D generation quality.}}




\begin{table}[!htb]

    
    \begin{minipage}{.5\linewidth}
          \centering
        \caption{\textbf{Ablation} on runtime}
        \label{tab:runtime}
        \vspace{-0.15in}
        \small
        \cready{
        \begin{tabular}{c|cc}
\hline
Methods       & FVD $\downarrow$          & Time$\downarrow$                   \\ \hline
B-I        & 992          & 2.25 hr           \\
B-II         & 729          & 1.10 hr                  \\ 
B-III         & 811          & 1.20 hr                  \\\hline
\rowcolor[HTML]{CBCEFB} 
Ours & \textbf{679} & \bf 35 min \\ \hline
\end{tabular}
     }
    \end{minipage}%
    \begin{minipage}{.5\linewidth}
 \caption{\bf User study}
 \vspace{-0.15in}
 \label{tab:user}
      \small
      \centering
      \cready{
\begin{tabular}{c|c}
\hline
Methods & Gen. Quality $\downarrow$ \\ \hline
B-I    & 3.14                 \\
B-II   & 2.35                \\ 
B-III  & 2.85                   \\
\hline
\rowcolor[HTML]{CBCEFB} 
Ours    & \bf 1.46                 \\ \hline
\end{tabular}
}
    \end{minipage} 
\end{table}

\begin{figure}
    \centering
    \includegraphics[width=1\linewidth]{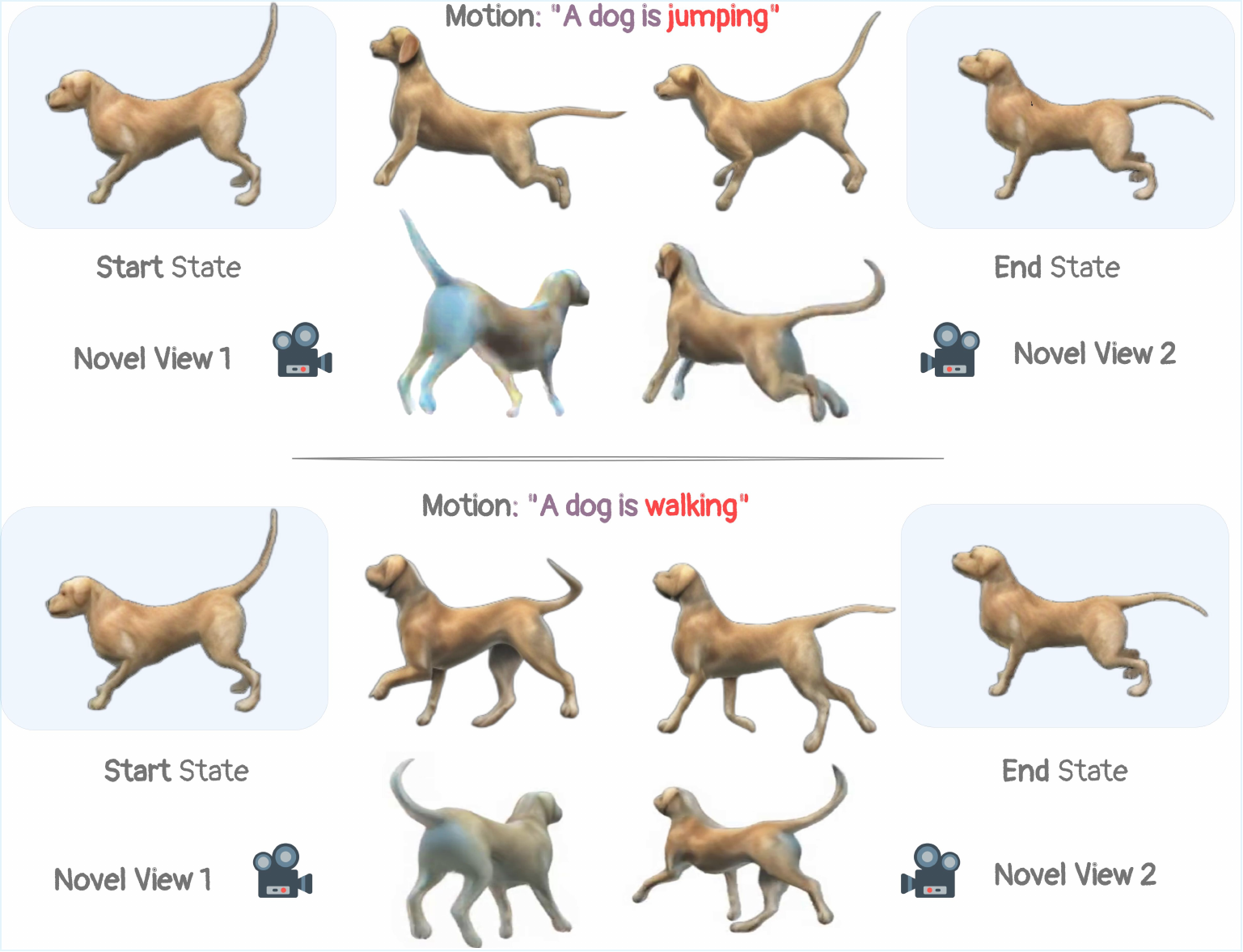}
    \caption{\textbf{Controllable Motions}. In-2-4D allows generation of diverse motions for the same start and end states}
    \label{fig:appl}
    \vspace{-0.2in}
\end{figure}

\noindent \textbf{Application: Customized 4D Motion.}
\saura{In contrast to most existing 4D generation methods \cite{ren2023dreamgaussian4d,zeng2024stag4d} that depend on SDS \cite{poole2022dreamfusion}, our approach improves controllability and motion diversity. While BLIP \cite{li2022blip} is used by default to extract motion prompts, users can input custom prompts to generate 4D motions for the same initial and final states. As shown in Fig.~\ref{fig:appl}, both \textit{jumping} and \textit{walking} motions of a dog are synthesized under identical start and end conditions. Despite motion complexity, our bottom-up 3D optimization ensures artifact-free novel view generation.}


%% file: main/05_conclusion.tex
\section{Conclusion, Limitations, Future Work}

 We introduce the novel task of generative 4D inbetweening from two single view images at distinct motion states. To address this challenging task, we leverage the capabilities of foundational video diffusion models to extract motion in between the states. We identify complex and large motions and divide them into fragments with simpler and smoother motions through a \emph{divide and conquer} approach. Using multi-view priors, we lift the object at different states to 3D and merge these simple 3D motions in a \emph{bottom-up} fashion with smoothness constraints into a flicker-free 4D motion. Although our work is able to outperform baselines but it is still a strong baseline on this challenging task and paves the way for further exploration and advancement.

Our method has some limitations. First, 
\saura{our method produces un-natural deformations when the in-between motion is extreme. Since the resulting videos are used to lift the object motion to the 3D space, the subtle movements may not look natural in 4D space.} A promising direction for future work would be to extend this approach to incorporate specific motion trajectories or other 2D or 3D conditional signals in 4D motion generation to provide more \saura{realistic dynamism}. Additionally, the 3D and 2D components do not currently interact in a way that allows mutual correction.
\cready{Finally, since our pipeline relies on an image-to-3D module, we also inherit its limitations.If the module fails to accurately reconstruct an object, the resulting 4D output will exhibit corresponding artifacts.}

\noindent \textbf{Acknowledgement} We thank all the anonymous reviewers for their insightful comments and constructive feedback. This work was supported by
the Natural Sciences and Engineering Research
Council of Canada (NSERC) Discovery Grant.